\documentclass[twocolumn,3p,final,times]{elsarticle}

\usepackage{epsfig}
\usepackage{amsmath,amssymb,amsfonts}
\usepackage{hyperref}
\usepackage{mathrsfs}
\usepackage{bbm}
\usepackage{slashed}
\usepackage{graphicx}
\usepackage{verbatim}

\newcommand{\be}{\begin{equation}}
\newcommand{\ee}{\end{equation}}
\newcommand{\bi}{\begin{itemize}}
\newcommand{\ei}{\end{itemize}}
\newcommand{\bea}{\begin{eqnarray}}
\newcommand{\eea}{\end{eqnarray}}
\newcommand{\bra}[1]{\langle\,#1\,|}          
\newcommand{\ket}[1]{|\,#1\,\rangle}          
\newcommand{\ud}{\mathrm{d}}		

\newcommand{\LCm}{{\scriptscriptstyle -}} 
\newcommand{\LCp}{{\scriptscriptstyle +}}

\newcommand{\LCperp}{{\scriptscriptstyle \perp}}

\newcommand{\bw}{\begin{widetext}}
\newcommand{\ew}{\end{widetext}}

\usepackage[usenames]{color}

\begin{document}

\begin{frontmatter}

\title{Radiation reaction in strong field QED}

\author{Anton Ilderton}
\ead{anton.ilderton@chalmers.se}

\author{Greger Torgrimsson}
\ead{greger.torgrimsson@chalmers.se}

\address{Department of Applied Physics, Chalmers University of Technology, SE-412 96 Gothenberg, Sweden}

\begin{abstract}
We derive radiation reaction from QED in a strong background field. We identify, in general, the diagrams and processes contributing to recoil effects in the average momentum of a scattered electron, using perturbation theory in the Furry picture: we work to lowest nontrivial order in $\alpha$. For the explicit example of scattering in a plane wave background, we compare QED with classical electrodynamics in the limit $\hbar \to 0$, finding agreement with the Lorentz-Abraham-Dirac and Landau-Lifshitz equations, and with Larmor's formula. The first quantum corrections are also presented.
\end{abstract}
\end{frontmatter}

%
%
Understanding radiation reaction presents one of the oldest problems in electrodynamics, and has recently seen a renewal of interest due to the potential impact, and detection, of recoil effects in high-intensity laser-matter interactions~\cite{DiPiazza:2011tq,Heinzl:2011ur}.  The years have seen many proposals for equations which describe a classical, radiating particle and avoid the problems of the Lorentz-Abraham-Dirac (`LAD') equation~\cite{LL-bok,MP,E,SOK-POP,deOca:2013qva}, but without consensus~\cite{O}.

For answers one would like to turn to (the classical limit of) QED, and the natural place to look for radiation reaction (`RR') is in photon emission from particles accelerated by background fields.  The potential use of intense lasers in measuring untested corners of QED (and physics beyond the standard model~\cite{Jaeckel:2010ni}) has lead to a great deal of activity in calculating QED processes in high-intensity laser backgrounds, an area called `strong field QED'.  Within this field, the one-photon emission spectrum of an electron has indeed been considered many times, but always found to give, in the classical limit, the spectrum of a particle moving according to the Lorentz equation. No recoil effects are seen~\cite{Boca:2009zz,Heinzl:2009nd,DiPiazza:2011tq}. Older Hamiltonian and `in-in' calculations, on the other hand, imply that one-photon emission is the key contributor to RR~\cite{Krivitsky:1991vt,Galley:2010es}. This has lead to confusion over which Feynman diagrams do/do not contain RR, and it has even been claimed that QED and classical electrodynamics are not compatible~\cite{SOK-POP,Sokolov:2010jx}. For a recent review of the situation, see the comprehensive discussion in~\cite{DiPiazza:2011tq}.

Given the great interest in this topic, and in the possibility of measuring RR at laser facilities, we will in this paper try to provide much-needed answers to the following questions. 1) Why have previous investigations of the photon spectrum not seen RR? 2) Where is RR in the $S$-matrix? Which diagrams contribute? 3) How are scattering results connected to those using other approaches? 4) What is the classical limit of quantum RR effects in QED?

Rather than pursue an equation (for which see~\cite{Krivitsky:1991vt} for perturbation theory and~\cite{Johnson:2000qd} for quantum stochastic corrections to LAD), we will consider scattering in QED, explain in general how RR arises, and use an explicit example to illustrate. This example is scattering in a plane wave background. With this choice we can treat the Lorentz force exactly to obtain a clean separation of recoil and non-recoil effects and perform the entire calculation with a single approximation: the usual coupling expansion of QED. 

We begin by reviewing how and where RR arises classically. We then turn to QED and show how the electron's momentum can be expressed in terms of $S$-matrix elements. Thus we identify which diagrams contribute to quantum RR. We then compare classical RR with the $\hbar\to 0$ limit of QED. We discuss related approaches, experimental implications, and applications to numerical simulations of strong field QED, before concluding.
\section{Classical radiation reaction}
This first section is intended to be pedagogical, since understanding the physics here will help when comparing classical and quantum results, and is enough to answer question `1)' in the introduction. The LAD equation for a radiating particle with orbit $x^\mu(\tau)$, proper time $\tau$, is ($c=\epsilon_0=1$)
\be\label{LAD}
	m \ddot{x}_\mu = e F^\text{ext}_{\mu\nu}(x){\dot x}^\nu + \frac{2}{3}\frac{e^2}{4\pi}(\dddot{x}_\mu {\dot x}_\nu - {\dot x}_\mu \dddot{x}_\nu ){\dot x}^\nu \;,
\ee
where $F^\text{ext}$ is an external field \cite{Coleman}. Retaining only the first term on the right hand side of (\ref{LAD}) gives the Lorentz force equation, the second term describes RR. The runaway solutions of LAD can be avoided through asymptotic boundary conditions~\cite{Coleman}, performing a reduction of order to obtain the Landau Lifshitz (LL) equation \cite{LL-bok},  using an alternate equation \cite{E,MP,SOK-POP}, or (since runaways are nonperturbative in the classical electron radius $r_0= e^2/4\pi m$) simply by expanding the orbit in $r_0$ (but see \cite{Zhang}).

To compare directly with QED, it is helpful to go back to the classical equations of motion and solve them perturbatively (rather than eliminating the radiation field to obtain LAD), in such a way that RR effects appear as corrections to the Lorentz force. To do so, define $f:=eF^\text{ext}$, let $F$ be the dynamical electromagnetic field with potential $A$, and $j$ the current. The classical equations of motion are then, schematically~\cite{Coleman},
\be\label{to-solve}
	m\ddot{x} = f \dot{x} + e F\dot{x} \;, \qquad \square A = ej \;.
\ee
(We assume $A_\text{ext}$ obeys Maxwell's equations in vacuum, $\square A_\text{ext}=0$). We solve (\ref{to-solve}) perturbatively, expanding\footnote{The UV divergence which arises in the derivation of LAD is proportional to the acceleration, and is usually removed by renormalising the particle mass~\cite{Coleman}, but see also~\cite{Gralla:2009md}. We will discuss this again when we come to the quantum theory.}
\be\begin{split}
	x &= x_0 + e x_1 +  \ldots \implies j = j_0 + e j_1 +  \ldots \\
	A &= A_0 + eA_1 +  \ldots \implies F = F_0 + e F_1 + \ldots
\end{split}
\ee
To zeroth order, we have $\square A_0 = 0$ and $\ddot x_0 = f(x_0) \dot x_0$, the Lorentz force equation. So, the particle moves according to the Lorentz force but, with appropriate initial conditions, $A_0=F_0=0$ and there is no radiation. To first order in $e$, we find a {\it homogeneous} equation for $x_1$. Since initial conditions can be fulfilled by $x_0$, we can set $x_1$ to zero, and the particle's motion is unaffected. At this order we also have $\square A_1 = j_0$,  and hence a nonzero radiation field $F_1$ sourced by $x_0$, i.e.\ by a particle moving under the Lorentz force. Radiation is therefore created at order $e$, but field observables are typically of order $e^2$, so the lowest order radiated energy, for example, is $E^2 +B^2 \sim e^2 j^2_0$, which does not contain recoil.

At second order in $e$, one finds $\square A_2 = F_2 = 0$ and  
\be\label{dethar}
	m\ddot x_2 =  f(x_0)\dot x_2 + x_2 \partial f(x_0)\dot{x}_0 +F_1(x_0)\dot x_0\;.
\ee
This {\it inhomogeous} equation yields a nonzero $x_2$. The particle's orbit is corrected due to the term $F_1\dot{x}_0$, i.e.\ by the fact that the particle has emitted the radiation $F_1$; this is radiation reaction. It appears at order $e^2$ in the electron's motion, as expected.  At third order, one finds that $F_3\not=0$, sourced by the radiating particle, i.e. $x_2$ contributes to the radiation field. Hence recoil effects appear in the radiated energy first at order $e^4$, through the $F_1 F_3\sim j_0 j_2$ cross term.

%
%
\section{Radiation reaction from QED}\label{RR}
%
\paragraph{Classical to quantum} We have now seen that to zeroth order in $e$, a particle is accelerated by an external field but does not emit. To first order, the particle's emission is accounted for. To second order, the effect of this emission on the electron is accounted for. The observables at this order include the corrected electron orbit, and the emission spectrum of a particle moving {\it according to the Lorentz equation}. At third order, the impact of recoil on the radiation field is accounted for, and enters the field's observables at order $e^4$.

What are the analogous results in the quantum theory? Consider QED with a background field. In the Furry picture~\cite{Furry},  interactions between quantised fields are treated perturbatively as normal, while the background is treated as a part of the `free' theory and, classically, affects the Lorentz force on the fermions \cite{Volkov:1935,Neville:1971uc,Dinu:2012tj}. Perturbative QED in the Furry picture therefore describes an expansion in powers of $\alpha = e^2/(4\pi\hbar)$ beyond a `free' theory without recoil, but with the Lorentz force. It is the quantum equivalent of the expansion used above. Thus, lowest order RR corrections should come from the classical limit of order $e^2$ Furry picture processes in QED, one of which is one-photon emission. However, previous investigations of this process (in a certain background, see below) found no RR effects~\cite{Boca:2009zz,Heinzl:2009nd,DiPiazza:2011tq}. We can solve this problem immediately; previous investigations have focussed on the emitted photon spectrum, but we saw above that RR effects are only visible at order $e^2$ in the electron sector. Because the probability of emitting a photon at all is already order $e^2$, recoil effects can only appear in the photon spectrum at order $e^4$ and higher, following e.g.\ multiple emissions~\cite{Seipt:2012tn,DiPiazza:2010mv,Neitz}. This solves problem 1) from the introduction. We expect, though, that one-photon emission should contribute to lowest order electron recoil. We will now see how.
\begin{figure*}[t!!!]
\includegraphics[width=0.2\textwidth]{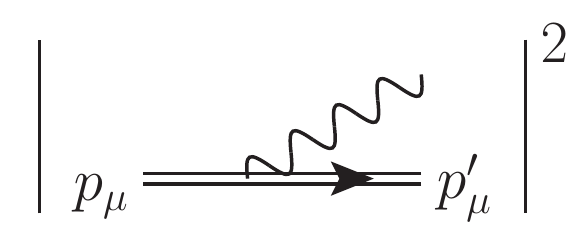}{\raisebox{5pt}{\Large ,}}\qquad \includegraphics[width=0.35\textwidth]{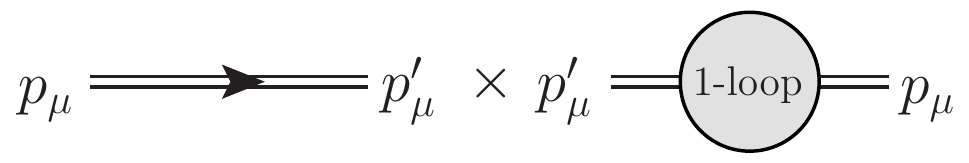}\hspace{5pt}\includegraphics[width=0.38\textwidth]{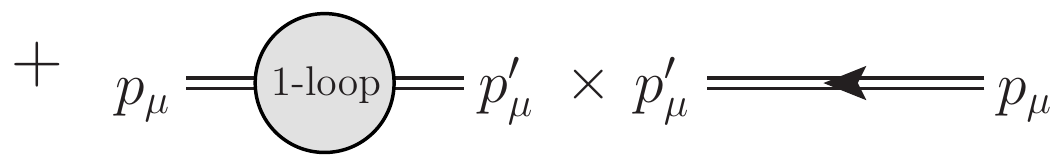}{\raisebox{5pt}{\Large .}}
\caption{\label{Diagrammen} Feynman diagrams contributing to radiation reaction at lowest order. A double line is the fermion propagator in a background field.}
\end{figure*}

\paragraph{RR from QED}
Consider a scattering experiment in which an electron is collided with a laser pulse, and its momentum is measured after exiting the beam. The classical theory, using (\ref{LAD}) or otherwise, predicts that the electron will have a certain momentum. The quantum theory predicts that, repeating the experiment, a distribution of final momenta will be measured. The expectation value of this distribution can be calculated in QED; it is the expectation value of the electron momentum operator in the final state of the scattering process. Expectation values (unlike probabilities or $S$-matrix elements) have natural classical limits; this is why approaches based on them, such as the in-in formalism~\cite{Johnson:2001rv}, are common in studying how classical RR arises from QED~\cite{Krivitsky:1991vt,Johnson:2000qd,Galley:2010es}. Such approaches are, though, not as widely used in particle physics as scattering amplitudes, so it is important to understand how RR arises in the latter.

Since experimental measurements are unlikely to be made within the fields of the laser, asymptotic results are sufficient. In this case, we can relate the expectation value of interest to familiar $S$-matrix elements and Feynman diagrams as follows. We begin with an incoming electron state $\ket{i}$, evolve it in time through a background field, and calculate the expectation value of the electron momentum operator, $P_\mu$, in the evolved state. If $S$ is the $S$-matrix, the expectation value is $\langle P_\mu \rangle = \bra{i}S^\dagger {P}_\mu S\ket{i}$. Inserting a complete set of asymptotic out states $\ket{f}$ with  $P\ket{f}=p^f\ket{f}$, we can write $\langle P_\mu \rangle$ in terms of $S$-matrix elements $S_{fi}=\bra{f}S\ket{i}$ as
\be\label{ev}
	\langle P_\mu \rangle =\sum\limits_f p^f_\mu\ |S_{fi}|^2  \;.
\ee
The sum in (\ref{ev}) runs over all Fock-sectors of free particle states, and can be interpreted as a sum of amplitudes for the processes $\mathrm{e}^\LCm \to \mathrm{e}^\LCm$+ anything. It was noted in~\cite{DiPiazza:2011tq} that obtaining the QED equivalent of LAD would require knowledge of the $S$-matrix at all orders; indeed, we see that this would be needed to perform the sum~(\ref{ev}) exactly\footnote{In fact, the situation may be even `worse': one may need more than the $S$-matrix, since finite time effects bring in new terms~\cite{US2}.}. To show that RR effects appear in QED at all, though, we need only show that $\langle P\rangle-\hat\pi\not=0$, where $\hat{\pi}_\mu$ is the final Lorentz force momentum of a particle after passing through the background field.

So far we have made no approximation, but to investigate in detail, it is simplest to use an expansion in a suitable parameter. We choose the coupling, since i) this is the most common approach to QED, ii) we do not need to assume anything about the background field or kinematic regime and iii) it gives us, at first nontrivial order, direct access to one-photon emission, the role of which, recall the discussion in the introduction, needs to be identified. (Of course, other expansions will be more suitable in other situations, especially for deriving phenomenologically useful results, see for example~\cite{DiPiazza:2010mv}.)

\begin{figure*}[t!!!]
\includegraphics[width=0.2\textwidth]{ett.pdf}{\raisebox{5pt}{\Large ,}}\qquad \includegraphics[width=0.35\textwidth]{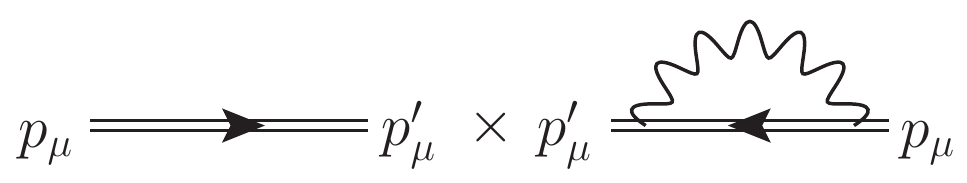}\includegraphics[width=0.38\textwidth]{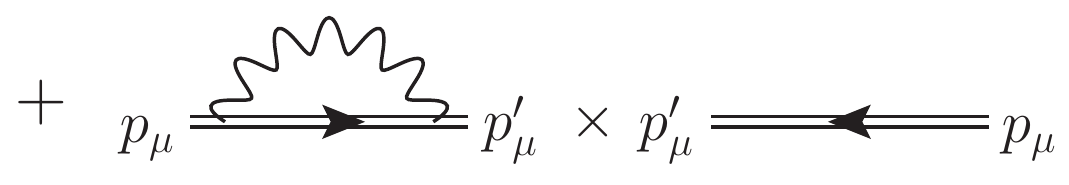}{\raisebox{5pt}{\LARGE .}}
\caption{\label{Diagrammen2} Feynman diagrams contributing to radiation reaction at lowest order, in a plane-wave background. In this case the tadpole vanishes.}
\end{figure*}

Each differential probability $|S_{fi}|^2$ in (\ref{ev}) has a Furry-Feynman diagram expansion in powers of $e^2$. (How many terms in this expansion one should consider to provide reliable predictions is a phenomenological question, which we will address elsewhere.) The zeroth order contribution to (\ref{ev}) comes from the one-electron sector, and is the process ${\mathrm e}^-\to {\mathrm e}^-$ without emission, at tree level. Classically, it describes only the effect of the Lorentz force. The next contributions to $\langle P_\mu\rangle$ come at order $e^2$ from two processes, see Fig.~\ref{Diagrammen}. First, photon emission at tree level, mod-squared. Second, the cross term of ${\mathrm e}^-\to {\mathrm e}^-$ to one loop, describing the self-field of the electron. Both processes are reminiscent of classical RR, and their sum gives (as we will shortly see explicitly) a finite, nonzero contribution to the final electron momentum, and to $\langle P_\mu\rangle$. This is lowest order quantum radiation reaction. We have answered questions 2) and 3) in the introduction.

It was noted in~\cite{Higuchi:2004pr}, see also~\cite{Hol} and below, that loop terms could be relevant in the classical limit. The loop contribution has not been considered in previous investigations of quantum RR within strong field QED. There is at least one good reason why it cannot be dropped; without the loops, $\langle P\rangle$ is in general infra-red divergent. However, it is well known that the inclusive sum of the loop and emission processes is IR finite and observable~\cite{Yennie:1961ad}; such sums are automatically included in (\ref{ev}) because $\langle P \rangle$ is an inclusive observable. This is an advantage of considering expectation values rather than individual diagrams or processes. (See~\cite{Gelis:2013oca} for related comments on Schwinger pair production.)

\section{Explicit example: plane waves}\label{exempel}
%

We now illustrate the classical and quantum discussions above, specifying to a plane wave background depending on invariant phase $\phi:=kx$ with $k^2=0$. We have $k_\mu=\omega n_\mu$ for $\omega$ an inverse-length scale, say a central frequency, and we can choose coordinates such that $kx = \omega x^\LCp$, lightfront time.  The transverse vector $a'_\LCperp$ ($ka'=0$) is the normalised electric field, $a'_\LCperp\equiv e E_\LCperp/\omega$. We consider pulses, so that $E_\LCperp(\phi)$ is either nonzero only in a finite $\phi$-range, or vanishes asymptotically, but is otherwise arbitrary.
\paragraph{Classical} The momentum $\pi^\mu:=m\dot x_0^\mu$ of a particle moving under the Lorentz force, initial momentum $p_\mu$, is~\cite{Hartemann,Dinu:2012tj},
\be\label{pi-lorentz}
	\pi_\mu(\phi) = p_\mu - a_\mu(\phi) + \frac{2pa(\phi)-a^2(\phi)}{2kp}k_\mu \;,
\ee
and one has that $kp$ is conserved, hence $\phi \propto \tau$. (From here on, an integral without variable is over lightfront time $\ud\phi$, $\pi\equiv \pi(\phi)$ and $\hat{\pi}\equiv \pi(\infty)$ is the final momentum as given by the Lorentz force in (\ref{pi-lorentz}).)  We proceed to~$x^\mu_2$, as introduced above. LAD, LL and the equations in~\cite{E,SOK-POP,MP}, while giving different results for {\it $x^\mu_2$} within the background, all agree on the first RR correction to the {\it final} momentum of a particle passing through the whole pulse\footnote{An explanation of why all the equations agree on this asymptotic result, to this order, is as follows. The first term in (\ref{klassiskt-ut}) can also be obtained by inserting the Lorentz orbit $x_\mu^0$ into Larmor's formula for the total radiated momentum. The second term in (\ref{klassiskt-ut}), representing momentum taken from the background, then follows by momentum conservation.}; this is $\delta\pi_\mu$, with
\be\label{klassiskt-ut}
	\delta\pi_\mu = \frac{2}{3}\frac{e^2}{4\pi}\frac{kp}{m^4}\int a^{\prime\, 2} \bigg(\pi_\mu-\frac{\pi\hat\pi}{kp}k_\mu\bigg) \;.
\ee
\paragraph{Quantum} We turn to the quantum calculation. We calculate the expectation value $\langle P\rangle$ to first nontrivial order, take the classical limit by taking $\hbar\to 0$ and compare with (\ref{klassiskt-ut}).  As above, the zeroth order contribution to $\langle P_\mu\rangle$ comes from tree-level scattering without emission, and yields $\langle P_\mu\rangle= \hat\pi_\mu$~\cite{Ilderton:2012qe}, the Lorentz force result. We proceed to order $e^2$.  Photon emission $\mathrm{e}^\LCm(p)\to \mathrm{e}^\LCm(p')+\gamma(k')$ in a plane wave, called nonlinear Compton scattering, yields an outgoing momentum $p^f = p'$ where
\be\label{vek1}
	p'_\mu := {\hat\pi}_\mu - k_\mu' + \frac{k'\hat\pi}{k(p- k')} k_\mu\;,
\ee
and this should be inserted into the integrated probability of emission to obtain its contribution to $\langle P_\mu\rangle$, see~\cite{Dinu:2012tj}. A large-(lightfront)-time divergence appears in this derivation. It can be removed by gauge invariance~\cite{Ilderton:2010wr} or by external line renormalisation~\cite{Boca:2009zz} (also \cite{Dinu:2012tj,Seipt:2010ya,Mackenroth:2010jr}), following the standard method in perturbative QED~\cite[\S 9]{Mandl:1985bg}. It is shown in~\cite{US2} using lightfront perturbation theory at finite time, that this first order divergence multiplies the zeroth order (Lorentz) {\it velocity}, or momentum. The classical divergence mentioned earlier is also order $e^2$ at the level of the equations of motion, but it is proportional to the {\it acceleration} in the LAD equation. In the perturbative solution of LAD, though, the divergence first appears at order $e^2$ multiplying the Lorentz momentum, as in the quantum case. Thus we find the same divergence in the classical and quantum theories to this order. One difference is that, in QED, the divergence is removed by operator, rather than mass, renormalisation~\cite{Collins,Dresti:2013kya}; it would be interesting to compare this with the classical renormalisation in~\cite{Gralla:2009md}.

Using similar methods, one finds that the loop in Fig.~\ref{Diagrammen2} leads to the same contribution as emission, except for an overall minus sign and that the final momentum is $p^f=\hat\pi$ rather than (\ref{vek1}). The total contribution to the final electron momentum at order $e^2$, call it $\delta\langle P_\mu\rangle$, is then ($\pi_j = \pi(\phi_j)$, $a_j=a(\phi_j)$)
\begin{align}\label{KVANT-SD}
&\delta\langle P_\mu\rangle =\displaystyle\frac{e^2}{4\pi\hbar}\int\limits_0^{p_\LCm}\!\frac{\ud k'_\LCm}{k'_\LCm}\!\int\!\frac{\ud^2k'_\LCperp}{(2\pi)^2} \frac{kp'}{kp}  \big(\hat{\pi}_\mu-p'_\mu\big) \\
&\nonumber \displaystyle \int\!\ud\phi_1\ud\phi_2 \exp\bigg[\frac{i}{\hbar}\int\limits_{\phi_1}^{\phi_2}\frac{k'\pi}{kp'}\bigg]\partial_2\partial_1 \bigg(\frac{m^2-g(\tfrac{kk'}{kp'})[a_2-a_1]^2}{k'\pi_2 k'\pi_1} \bigg),
\end{align}
and spin effects appear in $g(u):=\frac{1+(1+u)^2}{4(1+u)}.$ (Set $g=1/2$ for scalar QED.)   The result (\ref{KVANT-SD}) is finite, nonzero and has support on the difference between the electron's momentum following photon emission, (\ref{vek1}), and the Lorentz force (no recoil) result, $\hat\pi$. It is due to combined photon emission and self-energy effects. This is quantum radiation reaction in a plane wave background, and is exact in all parameters at order~$e^2$.

To take the classical limit of $\delta\langle P_\mu\rangle$, we note that, in a plane wave, it is the emitted photon's longitudinal momentum $kk'$ which is important. This `breaks the symmetry' of motion associated with the Lorentz force~\cite{Harvey:2011dp}, namely the conservation of $kp$. This is a sign of RR, as is seen explicitly when one solves LL in a plane wave~\cite{Exact}. Hence, we will rewrite (\ref{KVANT-SD}) to highlight its dependence on $kk'$. We therefore define $r_\mu$ by $k'_\mu = (kk'/kp) r_\mu$ and change variables $k'_\LCperp\to r_\LCperp$.  Noting that photon momentum has no classical analogue, but is equal to $\hbar\, \times\!$ wavenumber, we introduce $t$, which is the following simple combination of final (scattered) momenta,
\be\label{t}
	\hbar t = \frac{kk'}{kp'} = \frac{kk'}{kp-kk'} \;.
\ee
Changing variable from $k'_\LCm$ to $t$ removes $\hbar$ from the exponent, and makes all dependencies on $\hbar$ manifest. The result is that (\ref{KVANT-SD}) is exactly equal to
\bea\label{KVANT-SD2}
\delta\langle P_\mu\rangle = \displaystyle\frac{e^2}{4\pi}\!\int\limits_0^{\infty}\!\frac{\ud t}{(1+\hbar t)^2}\!\int\!\frac{\ud^2r_\LCperp}{(2\pi)^2} \left[\frac{r_\mu}{1+\hbar t}-\frac{r\hat\pi}{kp}k_\mu\right]\ \ \ \  \\
\nonumber	\displaystyle\int\! \ud \phi_1\ud\phi_2 \exp\bigg[ it \! \int\limits_{\phi_1}^{\phi_2} \frac{r\pi}{kp} \bigg] \partial_2\partial_1 \bigg(\frac{m^2-g(\hbar t)[a_2-a_1]^2}{r\pi_2 r\pi_1} \bigg) \;.
\eea
$\delta\langle P_\mu\rangle$ depends on $\hbar$ only through the combination $1+\hbar t$, so the classical limit $\hbar\rightarrow 0$ can be taken simply by expanding these factors in (\ref{KVANT-SD2}). From (\ref{t}), this is equivalent to assuming that $kk'\ll kp'$ (implying $kk'\ll kp$), i.e.\ that the momentum carried away by the photon is small compared to that of the electron. The classical limit therefore corresponds to neglecting quantum effects associated to emission of high energy photons. It remains to evaluate the $t$ and $r_\LCperp$ integrals at $\hbar=0$. Noting that $\delta\langle P_\mu\rangle$ is real and symmetric in $\phi_1\leftrightarrow\phi_2$, the $t$ integral can be symmetrised and immediately gives a delta function for the two $\phi$-integrals,
\be\label{t-integralen}
	\frac{1}{2}\int\limits_{-\infty}^\infty\!\ud t \cos\bigg( t\int\limits_1^2 \frac{r\pi}{kp}\bigg) = \pi\frac{kp}{r\pi(\phi_2)}\delta(\phi_2-\phi_1) \;.
\ee
This is the statement that interference terms drop out and only the incoherent piece of the $\phi$--integrals in (\ref{KVANT-SD}) survives as $\hbar\to 0$. This means that while quantum RR~(\ref{KVANT-SD}) is coherent in (lightfront) time, coming from an $S$-matrix element squared, classical RR is incoherent. This is a simple aspect of the decoherence intrinsic to quantum-to-classical transitions~\cite{Zurek:2003zz}. Using (\ref{t-integralen}) to eliminate one of the $\phi$-integrals, the remaining $r_\LCperp$ integrals become elementary, and we obtain
\be\label{sista}
	\lim_{\hbar\to 0}\delta\langle P_\mu\rangle=\frac{2}{3}\frac{e^2}{4\pi}\frac{kp}{m^4}\int a^{\prime\, 2} \bigg(\pi_\mu-\frac{\pi\hat\pi}{kp}k_\mu\bigg) = \delta \pi_\mu\;,
\ee
recovering the classical result (\ref{klassiskt-ut}) directly from QED. Quantum corrections are easily calculated using the same method. (The momentum integrals in $\delta\langle P\rangle$ can actually be calculated analytically,~\cite{Dinu:2013hsd}.) The order $\hbar$ contribution to (\ref{KVANT-SD2}) vanishes, while to order $\hbar^2$ one finds, for the longitudinal component $k\langle P\rangle$ for example,
\begin{align}\label{hbar2}
	&\frac{k\langle P\rangle}{kp} = 1 +\frac{2}{3}\frac{e^2}{4\pi}\frac{kp}{m^4}\int a^{\prime\, 2} \\
\nonumber	&-\frac{2}{5}\frac{e^2\hbar^2}{4\pi}\frac{kp^3}{m^{10}}\int (70+20s)\, a^{\prime\, 4} - (22+5s)m^2 a^{\prime\prime\, 2} \;,
\end{align}
where $s=\{0,\tfrac{1}{2}\}$ is the spin of the particle. This exhibits both kinematic and spin corrections due to quantum effects. While the classical term in (\ref{hbar2}) is strictly negative, the quantum term is typically positive, thus giving competing effects. We have answered question 4) in the introduction. 

\paragraph{Discussion}  Closely related calculations appear in~\cite{Elkina:2010up}, which recovers the dominant RR term from QED at high energy, and in~\cite{Sokolov:2010jx}, which concludes that QED supports the classical equation in~\cite{SOK-POP}. In perturbation theory, $k\delta\pi$ from (\ref{klassiskt-ut}) can be obtained from ordinary Compton scattering if the incoming photon flux is equal to the plane wave's energy density $a'^2$~\cite{Hartemann}. That paper finds that quantum and LAD results differ at large recoil. We can understand this by observing that accounting for larger recoil requires retaining higher powers of $\hbar$ in the expansion of (\ref{KVANT-SD2}), and these are non-classical effects, see~(\ref{hbar2}). An entirely perturbative calculation of the expectation value considered here appears in~\cite{Krivitsky:1991vt}, the advantage being that perturbation theory permits the consideration of arbitrary background fields. While our Sect.~\ref{RR} is general, and neatly relates the expectation value to the Feynman diagrams contributing to RR, our explicit example in Sect.~\ref{exempel} applies only to a plane wave background. Nevertheless, this has the benefit of making the classical limit simple, and giving insight into the role of the loop diagram, to which we now return.

We see from (\ref{vek1}) and below it that the loop does not contribute directly to classical RR, because it cancels against a term coming from the emission diagram. However, the cancelling terms are each $\mathcal{O}(1/\hbar)$. The loop therefore removes both IR {\it and} $1/\hbar$ singularities, and without it we would not be able to take the classical limit due to having a $1/\hbar$ divergence. In the remaining terms, recoil due to photon emission is proportional to $\hbar$~\cite{DiPiazza:2011tq}, but there is nevertheless a surviving classical contribution due to a cancellation with the QED coupling $\alpha\sim 1/\hbar$ as $\hbar\to 0$. See~\cite{Hol,Brodsky:2010zk} for related discussions.

In \cite{Higuchi:2005an}, the position shift of a scattered particle was compared between quantum and classical theories, and agreement found. This calculation was also for `asymptotic' times, i.e.\ measurements were made outside of the pulse. We have calculated momenta in this paper, as these are the natural variables in which to discuss scattering, but position is discussed in \cite{US2}.

Finally, consider higher orders terms. The sum in (\ref{ev}) is incoherent in particle number, just as in~\cite{DiPiazza:2010mv}, but each process is coherent in time in general. $n$-photon emission contributes to RR at order $e^{2n}$, along with all other processes of the same order, e.g.\ pair production, loop corrections, counterterms and so on. In general, these cannot be neglected from the outset. First, for consistency. Second, they remove IR and UV divergences. Third (and related), unitarity is violated without them: it was correctly noted in~\cite{DiPiazza:2010mv} that exclusive photon emission probabilities in plane waves easily exceed unity, but that unitarity could be restored by using a re-normalisation of the probabilities, following~\cite{G}. Both problem and solution here are typical of the IR in QED~\cite{BN,LN,K,Yennie:1961ad}. IR divergences only arise in `probabilities' which are not actually observable, due to the presence of indistinguishable processes. (Such `probabilities' can even be finite, but exceed unity, as for a class of plane wave backgrounds~\cite{Dinu:2013hsd}.) Physical observables, however, are always automatically IR finite. To remove IR divergences in QED (to all orders) one can either sum over degenerate processes~\cite{BN,LN,Yennie:1961ad}, calculate with physical rather than free asymptotic states~\cite{Lavelle:2005bt,Bagan:1999jk}, or calculate inclusive rather than exclusive observables~\cite{Yennie:1961ad}. The latter solution is automatic in our approach, since $\langle P \rangle$ is an inclusive observable. The terms which would otherwise violate unitarity are removed by the consistent, automatic inclusion of, e.g.\, the above loop. See the classic paper~\cite{Brown:1952eu} for the case of Compton scattering in QED.

Having provided our answers to the RR problems in the introduction, we now outline some applications and extensions of our results. 
\paragraph{Experimental implications} Our results suggest that it may be easier to see RR effects in electron spectra, rather than in photon spectra. This is supported by recent numerical simulations~\cite{Thomas}. In particular, this means that multi-photon emission experiments are not, strictly, necessary to observe RR. One needs only collide an electron with a laser, and measure the momentum of the scattered electron. To lowest order, one retains all events in which either zero or one photons are emitted (the photon momentum itself is not required), and compares this with (\ref{KVANT-SD}) or the appropriate numerical extension to more realistic fields. Whether this method is more suitable for measuring quantum or classical RR will be discussed elsewhere~\cite{US3}.

\paragraph{Classical equations} Asymptotic results following from the $S$-matrix, as used here, are sufficient for comparison with experiments aiming to observe radiation reaction, since scattering products will be measured far from interaction volumes. Nevertheless, the extension of our calculation to finite time and higher orders can distinguish between, and therefore rule out, different classical equations which predict different motion within the background\footnote{In geometries more complicated than a plane wave, both classical orbits and dressed quantum states will have a dependence on e.g.\ impact parameters, not just initial momenta as in a plane wave.  To recover a particular classical orbit from QED, one must then begin with a wavepacket peaked in both momentum and position.}. Our results appear in~\cite{US2}, and address the incompatibility between quantum and classical electrodynamics claimed in~\cite{SOK-POP}.

\paragraph{Numerical simulations} The use of powerful, large scale numerical models in strong field QED is increasingly popular~\cite{DiPiazza:2011tq,Elkina:2010up,SOK-POP,Ridgers,Thomas}. These approaches do not represent a nonperturbative discretisation of QED (as in lattice gauge theory), but instead are based on addition of perturbative cross-sections (primarily nonlinear Compton and stimulated pair production) to classical PIC codes. Despite this shared base, the codes differ in many aspects, for example in their implementation of RR. Some claim that RR must be included classically, since it does not occur in the photon emission diagram. We have shown that such implementations can double count the (classical) RR contribution, for photon wavelengths which are resolved by the numerical code. Given that interest lies in regimes where recoil effects are important, this could potentially be a serious overcounting. Our result (\ref{KVANT-SD}) provides a new, fully quantum benchmark with which to test that numerical codes correctly reproduce RR effects. We are currently investigating this~\cite{US4}.

\section{Conclusions}
We have addressed several questions related to radiation reaction in strong field QED. We have identified the processes and diagrams contributing to radiation reaction at lowest nontrivial order in $\alpha$, i.e.\ in the usual Furry picture perturbative expansion of strong field QED. Specifying to a plane wave background, we were able to evaluate the diagrams exactly to this order, thus eliminating potential ambiguities, and recovered classical radiation reaction in the $\hbar\to 0$ limit.

In this approach, RR arises as a small effect, i.e.\ as a correction to a leading order term which is the Lorentz force contribution. Our approach can of course be extended to higher orders, but we note that, apart from a few exact solutions~\cite{Exact,Bulanov}, there does not seem to be a fully nonperturbative approach to RR available in strong field QED. A potential option for investigating nonperturbative QED is offered by real-time lattice approaches, see~\cite{Hebenstreit:2013qxa}. 

We thank V.~Dinu, C.~Harvey, T.~Heinzl and M.~Marklund for useful discussions. The authors are supported by the Swedish Research Council, contract 2011-4221.

\end{document}